\documentclass[aps,twocolumn,superscriptaddress,nofootinbib,showpacs]{revtex4}
\usepackage{exscale}
\usepackage{graphicx}
\usepackage{amsmath}
\usepackage{latexsym}
\usepackage{amsfonts}
\usepackage{amssymb}

\def\be{\begin{equation}}
\def\ee{\end{equation}}
\def\bea{\begin{eqnarray}}
\def\eea{\end{eqnarray}}
\def\bma{\begin{mathletters}}
\def\ema{\end{mathletters}}

\def\0{\overline{0}}

\def\q0{\underline{0}}

\def\one{\leavevmode\hbox{\small1\normalsize\kern-.33em1}}

\newcommand{\Ket}[1]{\ensuremath{|#1\rangle}}

\newcommand\st[1]{{\mathrm{#1}}}

\newcommand{\entsp}{\stackrel{\scriptscriptstyle\wedge}{=}}

\DeclareMathOperator{\Trace}{Tr}
\newcommand{\Tr}[1]{\ensuremath{\Trace\left(#1\right)}}

\DeclareMathOperator{\partialTrace}{Tr}
\newcommand{\pTr}[2]{\ensuremath{\partialTrace_{#1}\left(#2\right)}}

\newcounter{Anumctr}

\newcounter{Bnumctr}

\newcounter{Cnumctr}

\newcounter{alphnumctr}

\begin{document}

\title{Continuous Variable Solution for Byzantine Agreement}

\author{Rodion Neigovzen}
\affiliation{Institut f\"ur Theoretische Physik, Universit\"at
  Hannover, D-30167 Hannover,Germany.}
\author{Anna Sanpera}
\altaffiliation[Also at:]{ Instituci\'o Catalana de Recerca i Estudis
  Avan{\c c}ats.}
\affiliation{Institut f\"ur Theoretische Physik, Universit\"at
  Hannover, D-30167 Hannover,Germany.}
\affiliation{Grup de F\'isica Te\`orica, Universitat Aut\`onoma de Barcelona, E-08193 Bellaterra, Spain.}

\begin{abstract}
We demonstrate that the Byzantine Agreement Problem (BAP) in its weaker version with detectable broadcast, 
can be solved using continuous variables Gaussian states with Gaussian operations. 
The protocol uses genuine tripartite symmetric entanglement, but 
differs from protocols proposed for qutrits or qubits. 
Contrary to the quantum key distribution (QKD) which is possible with all Gaussian states, 
for the BAP entanglement is needed, but not all tripartite entangled 
symmetric states can be used to solve the problem.
\end{abstract}

\pacs{03.67.Dd, 03.65.Ud, 03.67.-a}

\maketitle

An algorithm is commonly defined as a set of rules for solving a problem in a finite number of steps. 
One of the aims of Quantum Information is to provide new protocols and algorithms which exploit 
quantum resources 
to find solutions to problems which either lack a solution using classical resources, 
or the solution is extremely hard to implement. The term ``Byzantine Agreement Problem'' 
was originally coined by L. Lamport in the context of computer science.
In a cryptographic context it refers to distributed protocols in which 
some of the participants might have malicious intentions and could try to sabotage the 
protocol inducing the honest parties to take contradictory actions between them. 
This problem is often reformulated in terms of a Byzantine army, where there is a 
lieutenant general who sends the order of attacking 
or retreating to each one of his lieutenants. Those can also communicate pairwise to reach a common 
decision concerning attacking or retreating, knowing that there might be traitors among them including the general. 
A traitor could create fake messages to try that different parts of the army attack while others retreat putting 
the army at a great disadvantage. The question is thus whether there exists a protocol among all 
the officials involved  that, after its termination, satisfies the following conditions: 
The lieutenant general sends an order to his  $N-1$ lieutenants such that: 
(i) All loyal lieutenants obey the same order,
(ii) If the lieutenant general is loyal, then every loyal lieutenant obeys the order he sends. 
Lamport, Shostak and Pease proved in \cite{lamport2} that if the participants only share
pairwise secure classical channels, then Byzantine Agreement or broadcast (BA) 
is only possible iff $t<n/3$, where $n$ is the number of players and 
$t$ the number of traitors among them. 
In \cite{fitzi2001}, Fitzi and coworkers 
introduced the important concept of detectable broadcast,
and found a solution using quantum resources (see also \cite{fitzi2}). 
Detectable broadcast is said to be achieved if the protocol satisfies the following conditions: 
(i) If no player is corrupted, then the protocol achieves broadcast, and 
(ii) If one or more players are corrupted, then either the protocol achieves Byzantine Agreement 
or all honest players abort the protocol. Thus, in a detectable broadcast protocol, 
cheaters can force the protocol to abort, i.e. no action is taken, but in such cases all honest players 
agree on aborting the protocol so contradictory actions between them are avoided. 

Here we investigate the possibility of solving detectable broadcast using
Gaussian states and performing only Gaussian operations. 
Quantum Information with Gaussian states has become recently a very active 
research area \cite{book} since 
(i) Gaussian states (e.g. photons in coherent, squezeed or thermal states) are often the 
main quantum resource available in current experiments, 
(ii) in spite of the fact that they are elements of a Hilbert space of infinite dimensions 
they have a very compact and easy mathematical representation and 
(iii) it is desirable to classify entangled Gaussian states according its 
performance to several quantum information tasks as well as exploiting 
the differences between discrete and continuous variables scenario.

The protocols using entanglement as a resource\cite{fitzi2001,cabello,zheng} 
are based on 3 differentiated steps: (i) distribution \& test of the quantum states, 
(ii) measurements of the distributed states to establish a primitive 
and (iii) broadcast protocol. 
The (iii) step is fundamentally classical, since it uses the outputs 
of the measures of the quantum states to simulate a particular random generator (primitive).
In the simplest case, in which only 3 parties are involved, traditionally denoted by $S$, the sender, 
and the receivers $R_0$ and $R_1$, and at most, only one is
a traitor, the desired primitive 
generates for every invocation, with uniform distribution, a random permutation on the elements $\{0,1,2\}$, i.e. $(x_S, x_{R_0}, x_{R_1}) \in \{(0,1,2),(0,2,1),(1,0,2),(1,2,0),(2,0,1), (2, 1, 0)\}$ so that  each player $i$ has an assigned output $x_i$. In this primitive, 
each player ignores how the other two values 
are assigned to the other players. 
Furthermore, nobody else (besides the parties) have access to the sequences.
Quantum entangled states are used in the protocol 
to distribute classical private random variables 
with an specific correlation between the players, 
in such a way that any malicious manipulation of the data can be detected by all honest parties allowing them to abort the protocol. In the discrete case such primitive can be implemented with e.g. Aharonov states $\Ket{\mathcal{A}} = \frac{1}{\sqrt{6}}( \Ket{0,1,2} + \Ket{1,2,0} + \Ket{2,0,1}-\Ket{0,2,1} - \Ket{1,0,2} - \Ket{2,1,0})$ which assures the distribution and test part to be secure~\cite{fitzi2001}. 
Exploiting the fact that whenever the three qutrits are all measured in the same basis then all three results are different, the players are left -after discarding all the states used for the testing part- with a sequence of $j=1,2...L$ outputs 
schematically represented by the table below. We denote this quantum primitive as Q-flip.
\begin{center}
\begin{tabular}{|c|c|c|c|c|c|c|c|c|c|c|}
\hline $j$ & 1 & 2 & 3 & 4 & 5 & 6 & 7 & 8 & 9 & $\ldots$\\ 
\hline
\hline $S$ & 2 & 0 & 0 & 1 & 2 & 1 & 0 & 2 & 1 & $\ldots$\\ 
\hline $R_0$ & 1 & 1 & 2 & 0 & 1 & 0 & 1 & 0 & 2 & $\ldots$\\  
\hline $R_1$ & 0 & 2 & 1 & 2 & 0 & 2 & 2 & 1 & 0 & $\ldots$\\  
\hline 
\end{tabular}
\end{center}
After accomplishing the distribution \& test part of the protocol, the sender $S$ will broadcast a bit $x\in \{0,1\}$ ($0\entsp$``attack'', $1\entsp$``retreat'') to the two receivers using classical secure channels. Following \cite{fitzi2001} broadcasting (step iii) proceeds as follows:\\
\noindent (iii.1): We denote by $x_i$ the bits received by $R_i$ (notice that if the sender is malicious, 
the broadcasts bits $x_i$ could be different). Each receiver $R_i$ demands to $S$ to send him the indices $j$ for which $S$ got the result $x_i$. Each player $R_i$ receives a set of indices $J_i$.\\
\noindent (iii.2): Each $R_i$ test consistency of his own data, i.e.
checks weather his output on the 
set of indices he receives ($J_i$) are  all of them different from $x_i$. If so the data  
is consistent and he settles his 
flag to $y_i=x_i$. otherwise his flag is settled to $y_i=\perp$.\\
\noindent (iii-3): $R_0$ and $R_1$ send their flags to each other. If both flags agree, the protocol terminates with all honest participants agreeing on $x$.\\
\noindent (iii-4): If $y_i=\perp$, then player $R_i$ knows that $S$ is dishonest, the other player is honest and accepts his flag.\\
\noindent (iii-5): If both $R_0$ and $R_1$ claim to have consistent data 
but $y_0\neq y_1$, player $R_1$ demands from 
$R_0$ to send him all the indices 
$k\in J_0$ for which $R_0$ has the results $1-y_0$. $R_1$ checks now that (i) all indices $k$ from $R_0$ 
are not in $J_1$ and (ii) the output $R_1$ obtains from indices $k$ correspond to the value 2. 
If this is the case, $R_1$ concludes that $R_0$ is honest and changes his flag to $y_0$. If not, 
$R_1$ knows that $R_0$ is dishonest and  he keeps his flag to $y_1$. 
Detectable broadcast is in this way achieved.\\
%
Let us now introduce, analogous to the Q-flip, a bit-value primitive denoted by W-Flip. 
W-Flip is a random generator that (for every invocation), with uniform distribution, 
generates a random permutation on the elements $\{1,0,0\}$, i.e. $(x_S, x_{R_0}, x_{R_1}) \in  \{(1, 0, 0), (0, 1, 0), (0, 0, 1)\}$ so that each player $i$ has an assigned opt $x_i$. 
Invoking  twice a W-Flip
permits to map bit values to the trit values of the Q-flip $\{0,1,2\}$ according to 
the following rule:$(1,0)\rightarrow$ {\bf{0}},
 $(0,1)\rightarrow ${\bf{1}}, $(0,0)\rightarrow$ {\bf{2}} 
plus an additional element ``Z'' $(1,1)\rightarrow$ {\bf{z}}.
\begin{center}
\begin{tabular}{|c||c|c|c|c|c|c||c|c|c|}
\hline $j$ & \multicolumn{6}{c||}{Q-Flip} & \multicolumn{3}{c|}{Z-Flip}\\
\hline
$S$&\textbf{0}&\textbf{0}&\textbf{1}&\textbf{1}&\textbf{2}&\textbf{2}&\textbf{z}&\textbf{2}&\textbf{2}\\
\hline
$R_0$&\textbf{1}&\textbf{2}&\textbf{0}&\textbf{2}&\textbf{0}&\textbf{1}&\textbf{2}&\textbf{z}&\textbf{2}\\
\hline
$R_1$&\textbf{2}&\textbf{1}&\textbf{2}&\textbf{0}&\textbf{1}&\textbf{0}&\textbf{2}&\textbf{2}&\textbf{z}\\
\hline
\end{tabular}
\end{center}
To implement the Q-Flip with Gaussian variables via the W-Flip primitive 
we choose a family of fully inseparable 3-mode Gaussian states, completely symmetric under exchange of the players, characterized by a single parameter $a$ with covariance matrix \cite{giedke}:
\begin{equation}
\gamma(a) =\left(
\begin{array}{cccccc}
a&0&c&0&c&0\\
0&b&0&-c&0&-c\\
c&0&a&0&c&0\\
0&-c&0&b&0&-c\\
c&0&c&0&a&0\\
0&-c&0&-c&0&b
\end{array}\right)
\end{equation}
where $b=\frac{1}{4}(5a-\sqrt{9a^2-8})$ and $c=\frac{1}{4}(a-\sqrt{9a^2-8})$. For $a>1$  
all NPT-criteria\cite{giedke} are satisfied, so the state has genuine 3 partite entanglement.
Now to transform quantum states into a sequence of correlated private 
classical trits between the 3 players, 
each player measures the quadratures of the electromagnetic field, e.g. position (momentum)\cite{navascues}. 
Denoting by $\hat{X}_{i}$ the corresponding operators, and by $\hat{x}_{i}$
the measurement results, the players communicate classically with each other and agree on only those values for which $\left|\hat{x}_{S}\right|=\left|\hat{x}_{R_0}\right|=\left|\hat{x}_{R_1}\right|=x_0$ with $x_0>0$. Every player $i$ maps $\hat{x}_i=+x_0$ to the logical bit $x_i=1$ and $\hat{x}_i=-x_0$ to the logical bit $x_i=0$. 
Notice that W-flip requires that the probabilities of the outputs fulfill:
(i) $p(100)=p(010)=p(001)=p$ and $\delta=p(else)\ll p$, (ii) $\tilde{p}\rightarrow\frac{1}{3}$ and $\tilde{\delta}(else)\rightarrow0$ 
where 
$\tilde{p}(x_Sx_{R_0}x_{R_1})=p(x_Sx_{R_0}x_{R_1})/(\sum p(x_Sx_{R_0}x_{R_1})))$
are the conditional probabilities.
Using for the representation of the Gaussian states their Wigner function:
\begin{equation}
\label{wignerfunc}
W\left(\xi\right)=\frac{1}{\pi^{n}\sqrt{\det{\gamma}}}
\exp{\left[ -(\xi-d)^\st{T}\gamma^{-1}(\xi-d)\right] },
\end{equation}
where $d$ is a $2n$ real vector and  $\gamma$ the covariance matrix 
probabilities are straightforwardly calculated: 
\begin{eqnarray}
p(x_Sx_{R_0}x_{R_1})&=&\Tr{\rho_M^{x_Sx_{R_0}x_{R_1}}\rho_a}\nonumber\\
&=&(2\pi)^n\int W_M^{x_Sx_{R_0}x_{R_1}}(\xi)W_a(\xi)d^{2n}\xi.
\label{overlap}
\end{eqnarray}
Here $\rho_M^{x_Sx_{R_0}x_{R_1}}=\rho_S^{x_S}\otimes\rho_{R_0}^{x_{R_0}}
\otimes\rho_{R_1}^{x_{R_1}}$ 
describes the disentangled 3 Gaussian modes 
obtained after each party has performed a measurement in its subsystem 
and $\rho_a$ is the initial 3 partite entangled state described by $\gamma(a)$ and $d$. 
Optimal results are achieved for a displacement vector 
$d^T_W=-\frac{x_0}{3}(1,0,1,0,1,0)$, and yield:
\begin{eqnarray}
\label{eq:pAbs}
&&\delta_{1}=p(000)=C(a,\sigma)\exp\left(-\frac{4}{3}\frac{x_0^2}{K_1}\right),\\
&&\delta_{2}=p(111)=C(a,\sigma)\exp\left(-\frac{16}{3}\frac{x_0^2}{K_1}\right),\nonumber\\
&&\delta_{3}=p(110)=p(101)=p(011)=C(a,\sigma)\times\nonumber\\
&&\exp\left( \frac{-4 x_0^2\left(\sigma^2+\frac{1}{4}
\left[5a-\sqrt{9a^2-8} \right]\right)}{K_1 K_2}\right)\nonumber\\
&&p=p(100)=p(010)=p(001)=C(a,\sigma)\exp\left(-\frac{8}{3}\frac{x_0^2}{K_2}\right)\nonumber
\end{eqnarray}
with coefficients $K_1=\sigma^2+\frac{1}{2}\left[3a-\sqrt{9a^2-8} \right]$,
$K_2=\sigma^2+\frac{1}{4}\left[3a+\sqrt{9a^2-8} \right]$,
and the prefactor reads:
\begin{eqnarray}
&C&(a,\sigma)=\left[ \det{\left( \frac{\gamma_{M}+\gamma(a)}{2}\right) }\right]
^{-\frac{1}{2}}\\
&=&\frac{8}{\left(a-c+\sigma^2 \right) \left( b+c+\frac{1}{\sigma^2}\right) \sqrt{\left(a+2c+\sigma^2 \right) \left( b-2c+\frac{1}{\sigma^2}\right) }}\nonumber
\end{eqnarray}
In the approximation $a\gg1$ it follows that:
\begin{equation}
\label{eq:approxG_a}
\tilde{p}=\frac{1}{3}-\frac{4}{9}k+O(k^2);\;\;\; \tilde{\delta}_i\rightarrow 0
\end{equation}
with $k=\exp{\left[-\frac{4}{3}(\frac{x_0}{\sigma})^2\right]}$. The probabilities depend from 
the parameters $a$, $x_0$ and $\sigma$, but there exists a large region in the parameter space for 
which  $\tilde{p}=\frac{1}{3}$ and $\tilde \delta_i\rightarrow 0$. 
However, this condition is violated for  $a\leq a_{min}=\frac{5\sqrt{2}}{6}$
indicating that not all pure 3-mode symmetric entangled Gaussian states can be used 
to implement the W-Flip. Finally, we 
remark here that implementing the Q-Flip primitive by selecting those values for which $\left|x_{A}\right|,\left|x_{B}\right|,\left|x_{C}\right|\in\{0,x_0\}$ so that each player will associate the logical trit to the results of their measurement according to the following rule $q_m = 2,1,0$ if $x_m=+x_0,0,-x_0$ is not possible. In such case, the absolute and conditional probabilities 
never fulfill the requirements of the Q-Flip primitive. 

We move now to the distribution \& test part of the protocol 
which represents the first step in the execution of the W-Flip and has only 
two possible outputs: global success or global failure. In the case of failure a player assumes that something went wrong during the execution of the protocol 
and, therefore, he aborts any further action. The players share pairwise 
secure classical channels and secure (noiseless) quantum channels. 
The basic ingredient to implement a secure distribution \& test part, 
is based in the possibility that each player can check correlations by asking the other participants 
to send to him different random subsets of their systems. 
In doing so, it is possible to detect manipulation
of the data on an statistical basis and to abort the protocol if necessary. 
Here we focus on demonstrating security and ignore efficiency questions\cite{rodo}. 
This part proceeds as follows:\\
\noindent (i.1) $R_1$ prepares a large number  
$m=1,2,\ldots,M$ of systems in state $\gamma(a)$ with $d_W$ and 
sends one subsystem to $S$ and another to $R_0$.\\
\noindent (i.2) $R_1$ chooses (randomly) two disjoint sets of indices 
$K_i$, $i\in \{S,R_0\}$ ($ K_S\cap K_{R_0}=\emptyset$) 
and send $K_i$ to player $i$. Player $i$ is asked to send his subsystems $m\in K_i$ 
to player $j$. For each $m\in K_i$, participant $j$ measures the two subsystems in his possession while $R_1$ measures his subsystem. 
After communication on their results 
over secure classical channels they agree on those indices ${\tilde K}_j\subseteq K_j$ for which $|x_j|=|x_{R_1}|=x_0$. 
$R_1$ and $j$ check now whether the correlations of the W-Flip occur: $\tilde p_{100}=\tilde p_{010}=\tilde p_{001}=\frac{1}{3}$, $\tilde p_{\textnormal{else}}=0$. If the test was successful, i.e. if the measurement results were consistent with the assumption that the states have been distributed correctly, the players $i\in \{j, R_1\}$ set the flag $f_{i}=1$, otherwise $f_{i}=0$. 
In an analogous way, the test is performed for $S$ and $R_0$.\\
\noindent (i.3) Players $S$, $R_0$ and $R_1$ send their flags to each other.
Every player who receives a flag ``$0$'', sets his flag also to ``$0$''. 
Every player with flag ``$0$'' aborts the protocol. 
Otherwise the execution of the protocol proceeds. 
This step terminates the distribution \& test part.

In the second phase of the protocol a selection of the distributed systems is chosen to establish the bit sequences which will be use to implement 
the W-Flip primitive. In this phase, again honest parties may abort
the protocol if malicious manipulations occurs.\\
\noindent (ii-1) The players $S$, $R_0$ and $R_1$ agree upon a set of systems 
$m\in \tilde{M} \subset {M}$, which have not been discarded 
during the distribution \& test part.\\
\noindent (ii-2) Player $S$ chooses (randomly) two disjoint sets 
of subsystems labeled by indices $L_S^i\subset\tilde{M}$ 
and after sending index set $L_S^i$ to player $i$ demands player 
$i$ to send his subsystems $m\in L_S^i$ to him. 
In each case, the (random) choice $L_S^i$ is secret to party $j$, i.e. player $R_1$  has no information about the set $L_S^{R_0}$.
 Analogously this step is performed for $R_0$ and $R_1$.\\
\noindent (ii-3) $S$ performs now measurements on subsystems 
$m\in \tilde{M} - [L_{S}^{R_0}+L_{S}^{R_1}]=\hat{M}_{S}$. 
The analogous step is performed by the other players $i$.\\
\noindent (ii-4): After measuring their whole sequences, 
player $i\in\{S,R_0,R_1\}$ now announces
publicly, for which $m\in\hat{M}_{i}$ he measured $\pm x_0$, represented
by an index set $\hat{M}^m_i$. The order in which the players announce their measurement 
results can be specified initially and based e.g. on a rotation principle.\\
\noindent (ii-5) Without loss of generality let us specify this protocol step for player $S$. 
From the following sets $L_{S}^{R_0}\cap\hat{M}^{m}_{R_1}=:U^{R_1}_S$ and 
$L_S^{R_1}\cap\hat{M}^{m}_{R_0}=:U^{R_0}_S $ 
let $\tilde{U}^{R_i}_{S}\subseteq U^{R_i}_{S}$ for $i\in\{0,1\}$ 
be the index set for which the player $S$ measured 
$\pm x_0$ twice. Analogously to the first phase of the protocol, player $S$ can test 
if the output of his measures agree with the correlations 
of a proper W-Flip ($p(00)=p(10)=p(01)=1/3$ and $p(11)=0$) 
If the test is successful, the player sets his flag $f_{S}=1$, otherwise $f_{S}=0$. 
Same procedure is analogously performed by $R_0$ and $R_1$.
From this step on, players deal exclusively with the outputs of their measures,i.e. classical instruments.\\
\noindent (ii-6) $S$ checks correlation on his outputs in the following set
$\hat{M}:=\hat{M}^{m}_{S}\cap\hat{M}^{m}_{R_0}\cap\hat{M}^{m}_{R_1}\subseteq\hat{M}^{m}_{S}$.
If the test is successful, $S$ sets his flag $f_{S}=1$, otherwise $f_{S}=0$. 
$R_0$ and $R_1$ do the equivalent step.\\
\noindent (ii-7): For a randomly chosen set $V^{S}\subset\hat{M}$ player $S$ demands from 
$R_0$ and $R_1$ their measurement results. $S$ tests this control sample for the assumed 
W-Flip hypothesis. If the test is successful, $S$ sets his flag $f_{S}=1$, otherwise $f_{S}=0$. 
$R_0$ ($R_1$) perform this step with a set $V^{R_0}\subset\hat{M}\backslash V^{S}$ 
($V^{R_1}\subset\hat{M}\backslash (V^{S}\cup V^{R_0})$) respectively.\\
\noindent (ii-8) Every player with a flag $0$ aborts the execution of the protocol. 
Otherwise the players agree upon a set $W:=\hat{M}\backslash(V^{S}\cup V^{R_0}\cup V^{R_1})$ 
as the result of the W-Flip invocation, which must contain an even number of elements $2L$ so that 
a sequence of $L$ Q-flips can be extracted.
Broadcasting, as described previously by steps (iii) can now be implemented.
It is straightforward to show \cite{us}, that the 
strategy implemented by (iii) is secure and, if there is enough statistics,
the occurrence of the Z-flip can be safely determined.
Finally, we have considered two different types of errors 
that can occur when implementing the W-flip with
Gaussian states. The first one is just the error on obtaining a false combination of outputs, 
i.e. $\delta_i$. This error will propagate in the Q and Z-flips so that the probability of finding
a combination which is not appropriate is bounded from above from $\eta=1-(3\tilde p)^2$. 
The second source of ``errors'' are malicious manipulations. 
In order to manipulate the measurement results of other players, player $i$ could shift 
the local component of the displacement vector of the distributed state 
using local transformations. Notice that a shift on $x_0$ is equivalent to a shift on one 
of the components of displacement vector. That means that both kind of manipulations 
result on  the same change of probabilities for a given output calculated by the overlap
(see Eq.\ref{overlap}). Parameterizing the shift in the displacement 
vector by the parameter $K$, $d_W^T \mapsto {(d')_W^T} = -\frac{x_0}{3}(1,0,1,0,K,0)$

Thus player $i$ could determine via subsequent communication with the other players (step ii.4) 
with certain probability, the occurrence of the outputs of the other players 
thereby gain additional information. In contrast, direct measurement of a local manipulation 
without classical communication between $i$ and $j$ is impossible. 
This can be seen by realizing that the partial trace
$\pTr{C}{\rho_{\left( \gamma,d'\right) }}=\int W_{AB}\left(\xi_{AB}\right)
=\pTr{C}{\rho_{\left(\gamma,d\right)}}$
with
\begin{equation}
\gamma_{AB}=\left( \begin{array}{cccc}
a & 0 & c & 0 \\ 
0 & b & 0 & -c \\ 
c & 0 & a & 0 \\ 
0 & -c & 0 & b
\end{array} \right)\quad\textnormal{and}\quad
d_{AB}=\left( \begin{array}{c}
x_1 \\ 
0 \\ 
x_1 \\ 
0
\end{array} \right).
\end{equation}
A possible strategy for the traitor could consist of the following points:
(i) Discrediting honest players by manipulating the displacement vector in such a way that  
 non consistent combinations appear, (ii) to hide successful measurements to the honest players
which result in combinations that might be disadvantageous.
However, one can detect the effects of such manipulations by the test steps proposed in 
(ii-5)-(ii.7)\cite{us}.

To summarize, we have proposed an algorithm to solve detectable broadcast with 
continuous variables, i.e. using Gaussian states and Gaussian operations only. 
To the best of our knowledge this is the first truly ``multipartite'' algorithm for 
continous variables. Moreover, it can be straightfowdarly adapted to the discrete case 
by using 3 partite W-states\cite{duer} which presently are already created in ion 
traps\cite{blatt}. Finally, we have shown that although entanglement is 
needed to solve the problem not all entangled symmetric states can be used for this purpose 
and we have found a lower bound on the amount of entanglement required\cite{adesso1}. 
   
A.S. acknowledges discussions about BAP  with 
M. Lewenstein and D. Bru{\ss}. This work has been supported by the 
Deutsche Forschungsgemeinschaft, the EU (projects RESQ and QUPRODIS), 
the Kompetenzenetzwerk ``Quanteninformationsverarbeitung'' and the Spanish 
MCYT BFM-2002-02588.

\end{document}